\documentclass[
showpacs,showkeys,preprintnumbers,aps,prd]{revtex4}

\usepackage{amsmath}
\usepackage{amssymb}
\usepackage{revsymb}
\usepackage{graphicx}
\usepackage{dcolumn}

\begin{document}

\title{Structure of meson-baryon interaction vertices}
\author{
T. Melde$^{(1)}$, 
L. Canton$^{(2)}$, and 
W. Plessas$^{(1)}$}
\affiliation{
$^{(1)}$Theoretische Physik, Institut f\"ur Physik, Karl-Franzens-Universit\"at,
Universit\"atsplatz 5, A-8010 Graz, Austria\\
$^{(2)}$Istituto Nazionale di Fisica Nucleare, Sezione di Padova,\\
Via F. Marzolo 8, I-35131 Padova, Italy\\}

\begin{abstract}
We present a microscopic derivation of the form factors of strong-interaction
$\pi NN$ and $\pi N\Delta$ vertices within a relativistic constituent quark model.
The results are compared with form factors from phenomenological meson-baryon
models and recent lattice QCD calculations. We give an analytical representation of
the vertex form factors suitable for applications in further studies of hadron
reactions.
\end{abstract}


\maketitle
Understanding the meson-baryon strong-interaction vertices has been a hard and
long-standing problem. Attempts to derive a microscopic explanation, desirably on the
grounds of QCD, have not yet led to conclusive results. The problem is of considerable
importance not only in particle but also in nuclear physics. Practically all
realistic meson-exchange $NN$ potentials, 3$N$ forces, and $\pi N$ dynamical models
rely on certain inputs for strong form factors. Mostly they have been based on phenomenological arguments and one has usually employed monopole or dipole
parametrizations with cut-off parameters fitted to experiment. Different
parametrizations have big influences, e.g., on meson-baryon dynamical models (see
Refs.~\cite{Gross:1992tj,Surya:1995ur,Pascalutsa:2000bs,Sato:1996gk,Polinder:2005sm,
Polinder:2005sn}), 
on $NN$ potentials often used in present-day nuclear calculations (e.g., the
Nijmegen~\cite{Nagels:1975fb,Maessen:1989sx,Nagels:1978sc},
Bonn~\cite{Machleidt:1987hj,Machleidt:1989tm,Machleidt:2000ge},
and Argonne~\cite{Wiringa:1994wb,Wiringa:1984tg}
potentials), and on 3$N$ forces, see, e.g., refs.~\cite{Saito:1993se,Saito:2000bg}.
Consequently, a microscopic derivation of the meson-baryon interaction
vertices constitutes an important problem and it has long and often been asked for
(see, e.g., Refs.~\cite{Schutz:1995dj,Matsuyama:2006rp}).

The uncertainty about the meson-baryon strong-interaction vertices has even been
increased by the recent advent of lattice QCD
calculations~\cite{Alexandrou:2007zz,Erkol:2008yj}. These works have led to results
different among each other and partly distinct from earlier lattice QCD
calculations by Liu {\it et al.}~\cite{Liu:1994dr,Liu:1998um}. Lattice QCD results are furthermore at variance with form factors adopted so far in relativistic models
of meson-baryon dynamics~\cite{Sato:1996gk,Polinder:2005sm,Polinder:2005sn}.

Here, we perform a microscopic derivation of the strong meson-baryon form factors
on the basis of a relativistic constituent quark model (RCQM). It is free of any
phenomenological input (fit parameters), and the form-factor dependence on the
relativistic four-momentum transfer $Q^2$ is directly predicted from the RCQM,
which has already been successful in reproducing the invariant mass spectrum of
baryons~\cite{Glozman:1998fs,Glozman:1998ag}
and the electroweak structure of the nucleons and other baryon ground
states~\cite{Wagenbrunn:2000es,Glozman:2001zc,Boffi:2001zb,Berger:2004yi,Melde:2007zz}.
The same RCQM has recently been employed in a covariant study of the mesonic decays
of baryon resonances~\cite{Melde:2005hy,Melde:2006yw,Sengl:2007yq}
leading to results for partial decay widths qualitatively rather
different from previous nonrelativistic or relativized studies. The systematics
found in the relativistic decay widths for all the $\pi$, $\eta$, and $K$ decay modes
has subsequently also led to a partly new classification of baryon resonances into
flavor multiplets~\cite{Melde:2007iu,Melde:2008yr}.

\begin{figure}
\begin{center}
{\Large (a)}\hspace{0.5cm}
\includegraphics[clip=,width=6.cm]{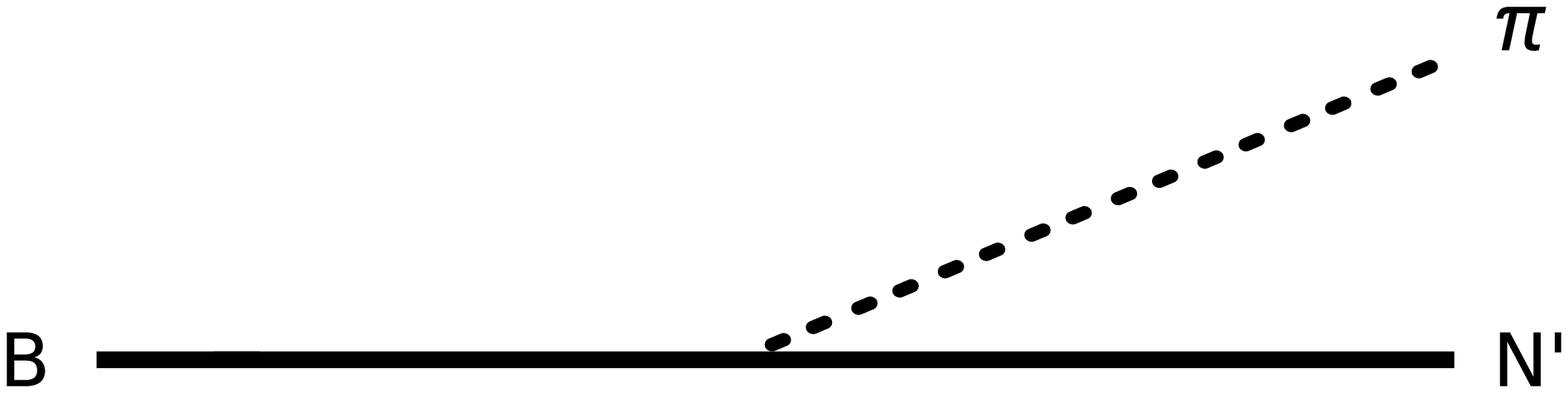}\\
\vspace{0.9cm}
{\Large (b)}\hspace{0.5cm}
\includegraphics[clip=,width=6.cm]{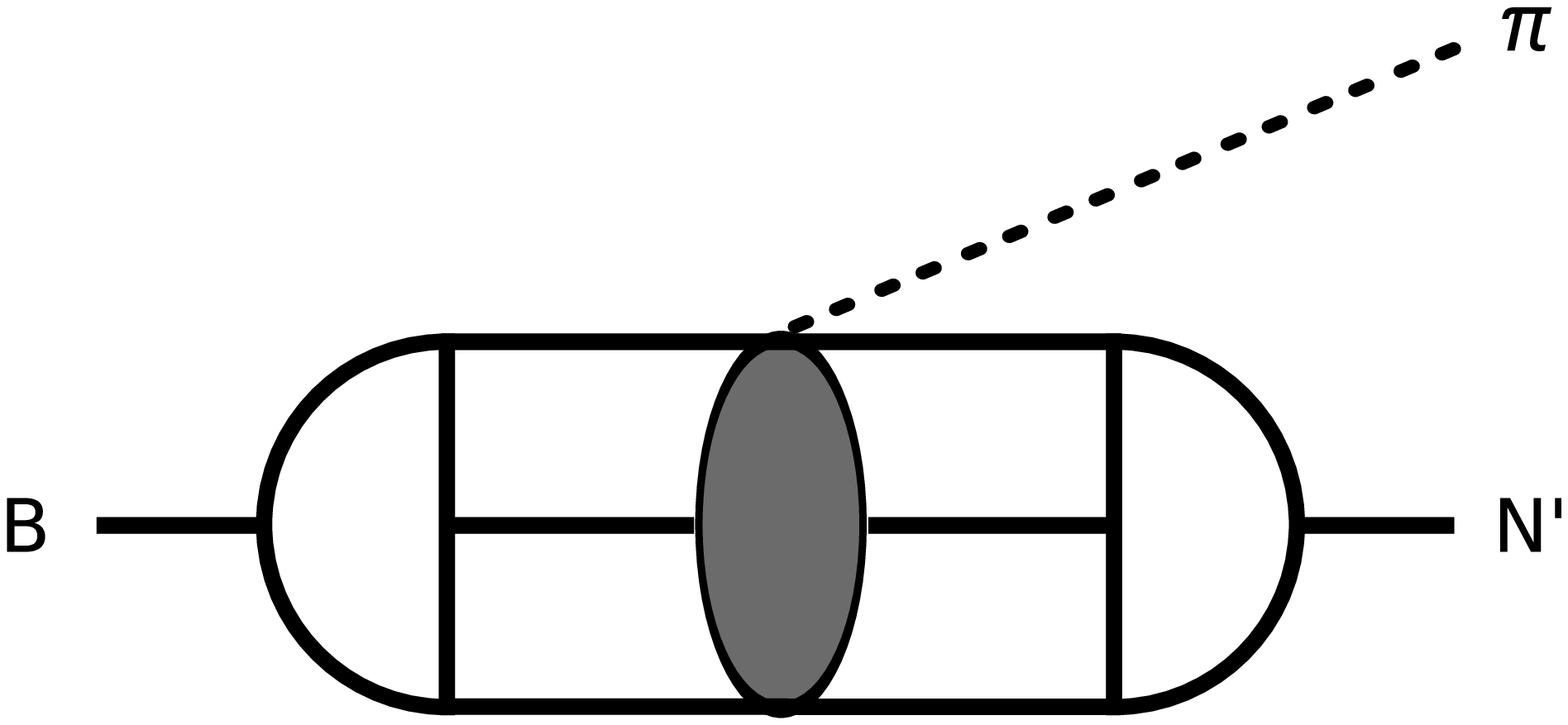}
\caption{\label{fig:hadron_3q} 
Graphical representation of the meson-baryon vertex (a) and the corresponding amplitude
in the RCQM (b).}
\end{center}
\end{figure}

In this paper we consider the $\pi NN$ and $\pi N\Delta$ form factors according
to the process depicted in Fig.~\ref{fig:hadron_3q}(a) and described by the matrix
elements of the hadronic interaction Lagrangian 
 \begin{equation}
 F_{i\rightarrow f}=\left(2\pi\right)^4\left< f\right|{\cal L}_{I}\left( 0\right)\left|i\right> \, ,
  \end{equation}
 where $f$ denotes the meson-emitting baryon and $i$ the final meson-nucleon state.
 The interaction Lagrangian densities  ${\cal L}_{I}$ are given by 
 \begin{eqnarray}
 {\cal L}_{I }^N&=&- \frac{f_{\pi NN}}{m_\pi}\bar \Psi\left(x\right)
 \gamma_5\gamma^\mu T\Psi\left(x\right)\partial_\mu\Phi\left(x\right)
 \\
  {\cal L}_{I }^\Delta&=&- \frac{f_{\pi N\Delta}}{m_\pi}
  \bar \Psi\left(x\right)T\Psi^\mu\left(x\right)\partial_\mu\Phi\left(x\right)+h.c.
 \end{eqnarray}
Herein, $\Psi$ is the nucleon Dirac field, $\Psi^\mu$ the $\Delta$ Rarita-Schwinger
field, $\Phi$ the meson field, and $T$ represents the transition operator for the
meson-emission process; $f_{\pi NN}$ and $f_{\pi N\Delta}$ are the $\pi NN$ and
$\pi N\Delta$ coupling constants, respectively.
We identify this process with the matrix elements of the (reduced) transition operator
${\hat D}_{\rm rd}^{\pi}$ for the same process sandwiched between eigenstates of the
invariant mass operator of the RCQM~\cite{Melde:2005hy}
 \begin{equation}
 F_{i\rightarrow f}^{\rm RCQM}=\left<V',M',J',\Sigma'\right| {\hat D}_{\rm rd}^{\pi}\left|V,M,J,\Sigma\right>\, ,
 \label{eq:trans_RCQM}
 \end{equation}
 graphically shown in Fig.~\ref{fig:hadron_3q}(b).
The baryon eigenstates in Eq.~(\ref{eq:trans_RCQM}) are characterized by the four-velocity
$V$, the invariant-mass eigenvalue $M$, and the intrinsic spin $J$ with z-component
$\Sigma$. 
We calculate the transition amplitude~(\ref{eq:trans_RCQM}) in the point form of
Poincar\'e-invariant quantum mechanics and take the transition operator according to the
spectator model~\cite{Melde:2004qu}, i.e.
\begin{multline}
\left<p'_1,p'_2,p'_3;\sigma'_1,\sigma'_2,\sigma'_3\right|
{\hat D}_{\rm rd}^{\pi}\left|p_1,p_2,p_3;\sigma_1,\sigma_2,\sigma_3\right>
=\\
3{\cal N}
\frac{i g_{qqm}}{2m_1\left(2\pi\right)^{\frac{3}{2}}}
{\bar u}\left(p'_1,\sigma'_1\right)
\gamma_5\gamma^\mu \lambda_m
u\left(p_1,\sigma_1\right)
\tilde Q_\mu
\\
2p_{20}\delta\left({\vec p}_2-{\vec p}'_2\right)
2p_{30}\delta\left({\vec p}_3-{\vec p}'_3\right)
 \delta_{\sigma_{2}\sigma'_{2}}
   \delta_{\sigma_{3}\sigma'_{3}}
\label{eq:hadrcurr}\, ,
\end{multline}
where the quark-meson coupling constant $g_{qqm}$ is fixed to the same value of
$\frac{g_{qqm}^2}{4\pi}=0.67$ as used in the Goldstone-boson-exchange (GBE)
RCQM~\cite{Glozman:1998ag,Glozman:1998fs}. 
The off-shell extrapolation of the transition amplitude is made by keeping all hadrons 
and quarks on their respective mass shells. Obviously it
implies energy non-conservation in the transition process. 
By virtue of the pseudovector-pseudoscalar equivalence the above construction also
guarantees that the pseudovector and pseudoscalar quark-meson couplings lead to the
same transition amplitude. 

As a function of the invariant four-momentum transfer squared in the space-like
region, $Q^2=-q^2>0$,
the strong $\pi NN$ form factor in the rest-frame of the meson-emitting baryon
is given by
\begin{equation}
\FL
G_{\pi NN}\left(Q^2\right)=\frac{1}{f_{\pi NN} }
\frac{m_\pi\sqrt{2\pi}}{\sqrt{2M_N}}
\frac{\sqrt{E'_N+M'_N}}{E'_N +M'_N+\omega}\frac{ F_{i\rightarrow f}^{\rm RCQM}}{Q_z} \,,
\label{eq:piNN}
\end{equation}
where the momentum transfer is taken into the $z$-direction.
Similarly, the $\pi N\Delta$ form factor reads
\begin{equation}
G_{\pi N\Delta}\left(Q^2\right)=-\frac{1}{f_{\pi N\Delta}}
\frac{3\sqrt{2\pi}}{2}\frac{m_\pi}{\sqrt{E'_N+M'_N}\sqrt{2M_\Delta}}
\frac{F_{i\rightarrow f}^{\rm RCQM}}{Q_z} \,.
\label{eq:piNDelta}
\end{equation}
Thereby we get the predictions of the RCQM for the $\pi NN$ and $\pi N\Delta$
coupling constants as well as the $Q^2$ dependences of the vertex form factors.

\begin{figure}
\begin{center}
\includegraphics[clip=,height=6.cm]{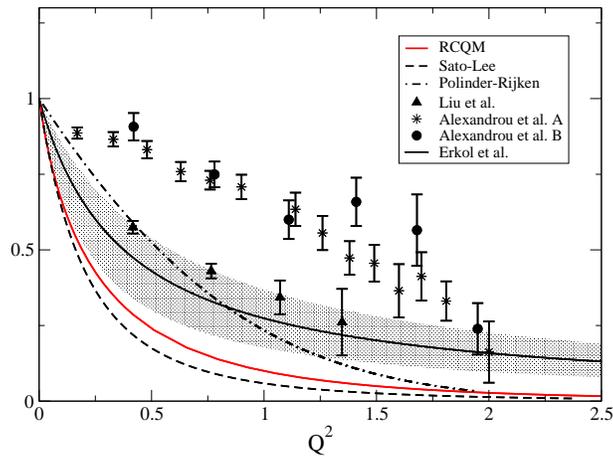}
\caption{\label{fig:Alex1} Prediction of the strong form factor $G_{\pi NN}$,
normalized to 1 at $Q^2=0$, by the
RCQM (solid/red line) in comparison to parametrizations from the dynamical meson-baryon
models of Sato-Lee~\cite{Sato:1996gk} and Polinder-Rijken~\cite{Polinder:2005sm,
Polinder:2005sn} 
as well as results from three lattice QCD
calculations~\cite{Alexandrou:2007zz,Liu:1994dr,Liu:1998um,Erkol:2008yj} (cf. the
legend); the shaded area around the result by Erkol {\it et al.} gives their theoretical
error band. See also the explanations in the text.}
\end{center}
\end{figure}

The $Q^2$ dependence of the $\pi NN$ strong form factor $G_{\pi NN}$ as predicted by
our RCQM is shown in Fig.~\ref{fig:Alex1}. There a comparison is made to parametrizations
from two dynamical meson-baryon models as well as results from lattice QCD calculations.
Our results compare best with the $\pi NN$ form factor of Sato-Lee~\cite{Sato:1996gk},
which represents a bare form factor, i.e. without hadron dressing. For comparison
we also give the dressed form factor from another dynamical meson-baryon model, namely
the one of Polinder-Rijken~\cite{Polinder:2005sm,Polinder:2005sn}. It exhibits a remarkably
slower fall off at small $Q^2$. The same is true for the lattice QCD results by
Liu {\it et al.}~\cite{Liu:1994dr,Liu:1998um} as well as the most recent ones
by Erkol {\it et al.}~\cite{Erkol:2008yj}. The latter essentially agree with each other,
where they both have made an extrapolation of their lattice data to the physical pion
mass or the chiral limit, respectively. The slowest fall off is shown by two types of
lattice QCD calculations
of Alexandrou {\it et al.}~\cite{Alexandrou:2007zz}, namely the one with quenched Wilson
fermions and a pion mass of 0.411 GeV (denoted as set A in Fig.~\ref{fig:Alex1})
and the one with dynamical Wilson fermions and a pion mass of 0.384 GeV (denoted as
set B in Fig.~\ref{fig:Alex1}); both are normalized using the coupling constant from
their linear fit. 

\begin{figure}
\begin{center}
\includegraphics[clip=,height=6.cm]{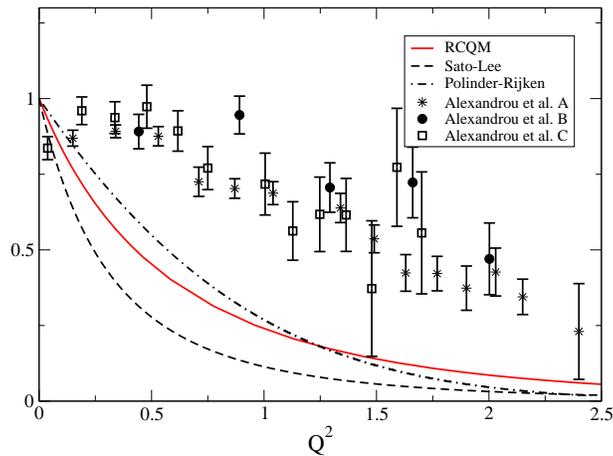}
\caption{\label{fig:Alex2} Same as Fig.~\ref{fig:Alex1} but for the
strong form factor $G_{\pi N\Delta}$.}
\end{center}
\end{figure}

The analogous predictions for the strong form factor $G_{\pi N\Delta}$ are given in
Fig.~\ref{fig:Alex2}. In this case there exist lattice QCD calculations only by
Alexandrou {\it et al.}~\cite{Alexandrou:2007zz}. In addition to the lattice data sets A
and B, as in Fig.~\ref{fig:Alex1}, we have added a further set denoted as C,
which corresponds to a calculation with a hybrid action and a pion mass of 0.353 GeV.
Again, the lattice QCD results by Alexandrou {\it et al.} are much above our predictions,
while our results now compare reasonably well with the parametrizations in both
dynamical models by Sato-Lee~\cite{Sato:1996gk} as well as
Polinder-Rijken~\cite{Polinder:2005sm,Polinder:2005sn}. We recall that the form factor
of the first corresponds to undressed hadrons whereas the one of the latter is dressed.

It is interesting to observe that the $Q^2$ dependence of both the $G_{\pi NN}$ and
$G_{\pi N\Delta}$ form factors resulting directly and in a parameter-free manner
from the RCQM qualitatively agrees with the parametrizations of the meson-baryon
vertices in the Sato-Lee model~\cite{Sato:1996gk}.
From their work the dressing effect is only visible
in the $\pi N\Delta$ case, where with increasing $Q^2$ the dressed form factor shows
a slightly faster fall off than the bare form factor.
There is no direct information on the behavior of undressed
form factors in the works by Polinder-Rijken~\cite{Polinder:2005sm,Polinder:2005sn}.
{\it A-priori} there
is no explicit dressing present in the RCQM yet. In principle, one should start out
with bare hadron masses in the construction of a constituent quark model. Then,
meson-cloud effects should be included explicitly. This is feasible now and remains as
a challenge for future constructions of RCQMs that aim to include the coupling to mesonic
channels. Thereby one could finally determine the dressing effects unambiguously
in the masses as well as in the form factors.

On the other hand, the strong form factors from the
lattice calculations show a (much) slower fall off with increasing $Q^2$. However,
even for the smaller differences between our results (as well as the bare form factors of
Sato-Lee) and the data sets by Liu {\it et al.} and Erkol {\it et al.} it
appears questionable that this would turn out merely as a dressing effect.
Regarding all of the
lattice data by Alexandrou {\it et al.} one has also to keep in mind that they correspond to
larger pion masses with no extrapolations applied. Thus it remains as an open question
if this is responsible for their rather weak $Q^2$ dependences. In any case
calculations/extrapolations towards smaller pion masses would be desirable.

The vertex form factors constitute an important input into all kind of
dynamical hadron models. Therefore we present the RCQM predictions shown above
in analytical forms suitable for further use. In particular, we adopt a form
intermediate between the usual monopole and dipole parametrizations
\begin{equation}
\label{a-pole}
G\left({\vec q}^{\,2}\right)=
\frac{1}{1+\left(\frac{\vec q}{\Lambda_1}\right)^2
+\left(\frac{\vec q}{\Lambda_2}\right)^4
}\, .
\end{equation}
This particular parametrization of the form factors depends on the three-momentum
transfer ${\vec q}^{\,2}$ rather than the four-momentum transfer $Q^2=-q^2$. It also
provides enough flexibility to represent our results above, while we could not obtain good
fits with an ansatz of either the standard monopole, dipole, or exponential type.
The monopole and dipole form factors are in fact contained in
our parametrization as limiting cases; Eq.~(\ref{a-pole}) reduces to the monopole form factor for $\Lambda_2 \rightarrow \infty$ and to the dipole form factor for
$\Lambda_1 =\Lambda_2 /\sqrt{2}$.

\renewcommand{\arraystretch}{1.4}
\begin{table}
\begin{center}
\caption{
Coupling constants and cut-off parameters of vertex form factors. The results of the RCQM
are represented according to the representation~(\ref{a-pole}) and are compared with the phenomenological models by Sato-Lee~\cite{Sato:1996gk} (SL) as well as
Polinder-Rijken~\cite{Polinder:2005sm,Polinder:2005sn} (PR). For the lattice QCD
calculations by Liu {\it et al.}~\cite{Liu:1994dr,Liu:1998um} (LIU),
Erkol {\it et al.}~\cite{Erkol:2008yj} (ERK) and set A of
Alexandrou {\it et al.}~\cite{Alexandrou:2007zz} (ALX) the monopole fit of
Eq.~(\ref{monopole}) is applied.
\label{tab:fit_parameter_vecq}
}
\vspace{0.2cm}
{\begin{tabular}{@{} lccc ccccc@{}}
& & RCQM & SL & PR$$ && LIU & ERK & ALX\\
\hline
&{\large$\frac{f_{N}^2}{4\pi}$ }& 0.0691 & 0.08 & 0.075 & & 0.0649 & 0.0481 & 0.0412  \\
$N$&$\Lambda_1$ & 0.451 & 0.453 & 0.940  & \phantom{ab} $\Lambda$ & 0.747  &  0.614  & 1.65 \\
&$\Lambda_2$ & 0.931 & 0.641 & 1.102 &  & - & - & - \\
\hline
&{\large$\frac{f_{\Delta}^2}{4\pi}$} & 0.188 & 0.334 & 0.478 & &  &   &  \\
$\Delta$&$\Lambda_1$ & 0.594 & 0.458 & 0.853 &   & &  &  \\
&$\Lambda_2$ & 0.998 & 0.648 & 1.014 &  & &  & \\
\hline
\end{tabular}}
\end{center}
\end{table}

The parametrization~(\ref{a-pole}) also allows to compare the form factor results on an
equal footing in terms of coupling constants and cut-off parameters. The corresponding
values are given in Table~\ref{tab:fit_parameter_vecq}. For the RCQM the size
of the coupling constant $\frac{f_{N}^2}{4\pi}$ (defined at $\vec q^2=0$)
for $\pi NN$ compares well with the phenomenological value
of approximately 0.075~\cite{Bugg:2004cm} as is the case for the Sato-Lee and
Polinder-Rijken models. The same is still true for the vertex form factors by
Liu {\it et al.}, whereas the more recent lattice QCD calculations by Erkol {\it et al.}
as well
as Alexandrou {\it et al.} yield much too small coupling constants. For the $\pi N\Delta$
coupling constant $\frac{f_{\Delta}^2}{4\pi}$ we obtain a smaller coupling constant
than Sato-Lee and
Polinder-Rijken. Due to the lack of firm experimental evidence it remains as an
open question, which is the most adequate value. In this context, it is noteworthy
that Polinder-Rijken found the bare coupling constant to be
$\frac{f_{\Delta}^2}{4\pi}=0.167$, i.e. a much smaller value than the dressed one
but closer to our RCQM prediction.  

For the $\vec q^{\,2}$ dependence of the vertex form factors the best fits are
obtained with the cut-off parameters in Table~\ref{tab:fit_parameter_vecq}. We
recall that the original results of Sato-Lee and Polinder-Rijken are given as dipole
and exponential form factors, respectively. For the lattice QCD
calculations by Liu {\it et al.} as well as Erkol {\it et al.} there exist
monopole form-factor fits in terms of $Q^2$
\begin{equation}
\label{monopole}
G\left(Q^{\,2}\right)=
\frac{1}{1+\left(\frac{Q}{\Lambda}\right)^2
}\, ,
\end{equation}
and the data from the linear fit by
Alexandrou {\it et al.} can also be cast into the same type of monopole representation.
As is seen from Table~\ref{tab:fit_parameter_vecq} the cut-off parameter $\Lambda$
relating to the form-factor results of Alexandrou {\it et al.} is remarkably high.

In summary, we have presented a parameter-free microscopic description of the strong
$\pi NN$ and $\pi N\Delta$ vertex form factors with a fully relativistic constituent
quark model. The $Q^2$ dependences of the RCQM form factors are qualitatively similar
to the ones parametrized along the phenomenological dynamical meson-baryon model
by Sato-Lee with bare hadrons. The RCQM predictions require a
parametrization intermediate between a monopole and dipole form. In particular,
this also suggests that there is no preference
for a cut-off of exponential type, which has sometimes been claimed to be
suggested from microscopic quark-model considerations.
In addition, our study reveals that the structure of the $\pi N\Delta$ vertex is
sizably different from the $\pi NN$ one, with cut-off parameters of up to 25\%
larger. This is at variance with form factor parametrizations often used in
phenomenological models, where the $\pi NN$ and $\pi N\Delta$ cut-offs are assumed
of similar size~\cite{Sato:1996gk,Polinder:2005sm,Polinder:2005sn} or even decreasing
in the transition from $\pi NN$ to $\pi N\Delta$.

The lattice results on the vertex form factors are qualitatively distinct from the
RCQM predictions and they also differ considerably among each other.
It remains to be shown if the slower fall-off
of the lattice QCD form factors with increasing $Q^2$ is just a dressing effect.
In this regard it appears most important to have on the one hand more lattice data
and on the other hand (relativistic) quark-model studies that can quantitatively
pin down the contributions of hadron dressing. 

\begin{acknowledgments}
This work was supported by the Austrian Science Fund (FWF Project P19035).
We thank C. Alexandrou and T.~T. Takahashi for useful communications regarding 
the representation of their lattice QCD data.
\end{acknowledgments}
\addcontentsline{toc}{chapter}{Bibliography}
%
%

%
\end{document}